\documentclass[prb,twocolumn,superscriptaddress,floatfix,showpacs]{revtex4}
\usepackage[dvips]{graphicx}
\usepackage{dcolumn}
\usepackage{amsmath}
\usepackage{amssymb}
\usepackage{amsbsy}
\begin{document}
\input{epsf}
\title{Effect of interfacial Cr on magnetoelectricity of 
 Fe$_2$/CrO$_2$/BaTiO$_3$(001)}
\author{M. H\"olzer}
\affiliation{Institut für Physik, Martin-Luther-Universit\"at Halle-Wittenberg,
 D-06099 Halle, Germany} 
  \author{M. Fechner}
\affiliation{Max-Planck-Institut f\"ur Mikrostrukturphysik, Weinberg
  2, D-06120 Halle (Saale), Germany}
  \author{S. Ostanin}
\affiliation{Max-Planck-Institut f\"ur Mikrostrukturphysik, Weinberg
  2, D-06120 Halle (Saale), Germany}
\author{I. Mertig}
\affiliation{Max-Planck-Institut f\"ur Mikrostrukturphysik, Weinberg
  2, D-06120 Halle (Saale), Germany}
\affiliation{Institut für Physik, Martin-Luther-Universit\"at Halle-Wittenberg,
 D-06099 Halle, Germany}
\date{\today}

\begin{abstract}
 On the basis of first-principles calculations we study the effect of interfacial 
 Cr on the magnetoelectric properties of a composite multiferroic
 Fe$_L$/BaTiO$_3$(001), with the Fe thickness $L\le$2 monolayers.
 The use of the CrO$_2$-terminated interface instead of TiO$_2$ may significantly 
 enhance magnetoelectricity in the system, showing an unexpected change in 
 magnetization induced by the electric polarization reversal. 
 In the case of $L=$2, for instance, the magnetic order of the Fe bilayer can be switched 
 from nearly zero ferrimagnetic to ferromagnetic upon polarization reversal.
\end{abstract}


\maketitle

\section{Introduction}

 The occurrence of ferroelectricity and ferromagnetism in the same phase of a 
 so called multiferroic \cite{WangLiuRen:2009} (MF) material allows both a 
 switchable electric polarization, $\bf P$, and  a switchable magnetization $\bf M$.
 More precisely, when an applied electric field displaces the magnetic ions of 
 the multiferroic this affects the magnetic exchange coupling or, vice versa, 
 the external magnetic field, $\bf H$, induces $P_i \sim {\alpha}_{ij} H_j$, 
 where ${\alpha}_{ij}$ is the magnetoelectric (ME) tensor and $(i,j)=x,y,z$. 
 When ${\alpha}$ is sufficiently strong this phenomenon may allow to store 
 information in nanometer-sized memories with four logic states 
 \cite{Eerenstein:2007p281,Cheong:2007p446,Zavaliche:2007p7964}.

 The classification of multiferroics is based on different mechanisms 
 of induced polarity \cite{Khomskii:2009p8591}. The type-I class of multiferroics 
 contains numerous perovskitelike materials in which $\bf P$ appears at higher 
 temperatures than magnetism. In these materials, $\bf P$ and $\bf M$ weakly 
 interact with each other and, therefore, ${\alpha}$ is marginal there. 
 In type-II MF, such as TbMnO$_3$, ferroelectricity is driven by the electronic 
 order degrees related to a spin-orbit mechanism in conjunction mostly with 
 the spin-spiral magnetic arrangement via the Dzyaloshinskii-Moria antisymmetric 
 exchange. The latter creates $\bf P \sim {\bf r}_{ij} \times [{\bf S}_i \times {\bf S}_j]$, 
 where ${\bf r}_{ij}$ is the vector connecting neighboring spins ${\bf S}_i$ and ${\bf S}_j$.
 Some of the type-II MFs may disclose a relatively large ME coupling. However,
 their ferroelectricity is caused by a particular type of magnetic order, which exists 
 only at low temperature and which is predominantly antiferromagnetic.

 Studies based on density functional theory (DFT) have significantly contributed to 
 this rapidly developing field of multiferroics \cite{PicozziEderer}. For instance,
 calculations from first-principles predict that the ME effect appears when a meV voltage 
 is applied across the interface between the two unlike terminations, such as 
 SrRuO$_3$/SrTiO$_3$ \cite{Rondinelli:2008p6059}. The interface ME effect might be 
 intrinsically enhanced by the use of material with high spin polarization. Indeed, 
 a more robust scenario of magnetoelectricity occurs in epitaxially grown two-phase MF 
 consisting of ferroelectric and ferromagnetic components.  
 {\it Ab initio} calculations suggest that chemical bonding at the Fe/BaTiO$_3$(001) 
 interface is the source of strong ME coupling \cite{Duan:2006p278,Fechner2}.
 Moreover, for the two opposite directions of $\bf P$ ($P_\downarrow$ and $P_\uparrow$), there are 
 rather noticeable differences of 0.1--0.2${\mu}_B$ in the magnetic moments of interfacial Fe and Ti.
 This is a very promising phenomenon, which is entirely confined to the ferroelectric/ferromagnetic
 interface. The interface ME effect \cite{Duan:2006p278} defines the change in ${\bf M}$ 
 at the coercive field $E_c$: 
\begin{equation}
 {\mu}_0 \Delta M \approx \alpha E_c.
\end{equation}
 For Fe/BaTiO$_3$(001), the estimated \cite{Fechner2} $\alpha$ of 
 $\sim$2$\times$10$^{-10}$ G cm$^2$/V is two orders of magnitude larger than that predicted 
 for SrRuO$_3$/SrTiO$_3$.

 Currently, {\it ab initio} calculations which explore the trends and basic physics of 
 magnetoelectrics, go ahead of experiment. For a single Fe monolayer (ML) on BaTiO$_3$(001), 
 DFT predicts that perpendicular anisotropy is favored to in-plane anisotropy by 
 0.7 meV (0.5 meV) per Fe atom for $P_\downarrow$ ($P_\uparrow$) \cite{Fechner2}. 
 Although the spin reorientation transition under switching of $\bf P$ is not found from first 
 principles, the ME coupling alters the magnetocrystalline anisotropy energy by $\sim$50\%. 
 The magnetic order of Fe/BaTiO$_3$ can be tuned by the Fe layer thickness to almost 
 zero-$\bf M$ ferrimagnetic upon deposition of a second Fe ML \cite{Fechner2}. 
 Ferromagnetic order is restored for the Fe films thicker than 3 ML where the shape 
 anisotropy energy favors in-plane alignment of $\bf M$ \cite{Duan:2008p6861}.
 Epitaxial growth of the two-phase MF thin films of high quality continues to be very challenging.
 A 30-nm thick Fe(001) film has been grown recently on a ferroelectric BaTiO$_3$(001)
 substrate \cite{Yu:2008p7600}. For this composite MF, the trends of magnetic anisotropy
 are in good agreement with the corresponding {\it ab initio} calculations
 \cite{Fechner2,Duan:2008p6861}.
 Until recently, the DFT studies of the interface ME coupling were focused on chemically 
 perfect films and superlattices with no impurities. Modeling the two different 
 Fe$_3$O$_4$/TiO$_2$/BaTiO$_3$(001) interfaces, within the DFT, Niranjan {\it et al.} 
 \cite{Niranjan:2008p7603} have found that ME coupling is stronger for the O-deficient 
 type of the Fe$_3$O$_4$ interface. Therefore, the presence of extra oxygen or oxygen 
 vacancies at the biferroic interface plays an important role. The effect of iron oxidation 
 on the ME coupling of Fe/$A$TiO$_3$(001) ($A$=Ba, Pb) was simulated \cite{Fechner-PRB-2009} 
 from first principles for oxygen coverages ranged between $0.5$ and $2.0$ adsorbed O 
 atom per Fe atom. The calculations suggest that the magnetic properties of the Fe monolayer 
 are gradually degraded with increasing O coverage. However, the change in magnetization 
 which is induced by the ${\bf P}$ reversal remains robust. Thus, the surface oxidation of 
 composite MFs cannot destroy their potentially switchable magnetoelectricity. 

 It is well known that both the magnetic order of Fe-films and the related magnetic anisotropy 
 are very sensitive to the presence of some other 3d elements. The alloying effect may result 
 in important changes in magnetoelectricity and therefore, the DFT based modelling
 of chemical order in composite multiferroics would be useful. The effect of
 Fe-Co alloying on magnetoelectricity of thin-film Fe/BaTiO$_3$(001) has been studied
 recently\cite{FeCo-on-BTO} from first principles using the coherent-potential 
 approximation to DFT. It was found that the presence of $>$0.25 Co at.\%\ per Fe atom
 stabilizes the ferromagnetic order in the two-ML thick and magnetically soft Fe-films. 
 In this work, we investigate the ME coupling in the 1-ML and 2-ML thick Fe on BaTiO$_3$(001) 
 (BTO), with a CrO$_2$ interfacial layer instead of TiO$_2$. 
 Chromium dioxide (CrO$_2$) is an experimentally proven half metal, which shows
 a Curie temperature of 392 K and which possesses 
 the largest spin polarization so far reported for this class of materials.
 As a consequence of the half-metallic feature of CrO$_2$ , the occupied Cr 3d bands 
 are fully spin polarized, leading to the spin moment of 2 ${\mu}_B$ per formula unit.
 Now we explore whether such a CrO$_2$-terminated interface of BTO  
 enhances the ME coupling in Fe$_L$/BTO(001).  
 In particular, for $L=2$ we observe a dramatic change of magnetization 
 in the topmost Fe ML under polarization reversal.  

\section{Method}

\begin{figure}
  \includegraphics[width=1.0\linewidth]{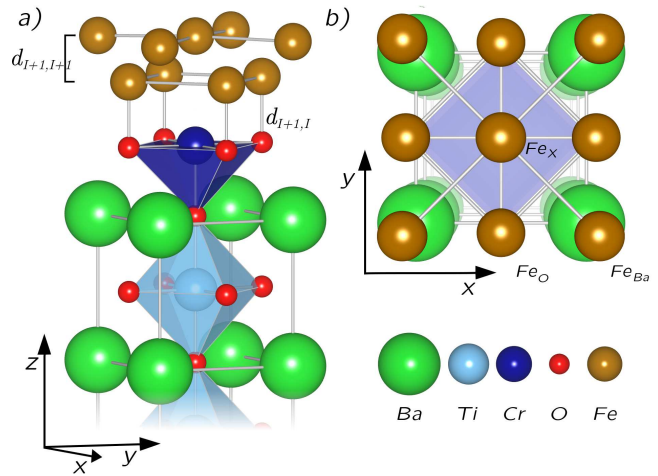}
  \caption{\label{surface_cell} Top layers of the (Fe$_2$)$_{L=2}$/CrO$_2$/BaTiO$_3$(001)
  unit cell are shown as side- and top view in the panels (a) and (b), respectively.
  In panel (b), the interfacial Fe atoms above O are marked as ``Fe$_O$"
  while the Fe atoms of the second ML, which sit above the perovskite cations are marked 
  with the labels ``Fe$_{Ba}$" and ``Fe$_{X}$".}   
\end{figure}
 
 To model the Fe$_{L}$/BaTiO$_3$(001) biferroic system within a 
 slab geometry we used a 5-unit-cell ($\sim$2-nm) thick BTO supercell covered by
 an Fe monolayer or Fe bilayer ($L =$1,2). A 2-nm-vacuum layer separates the slabs along [001].
 For tetragonal BTO the equilibrium lattice parameters $a=$3.943 \AA\ and
 $c/a=$1.013 were used. The Fe positions and atomic positions of the two top BTO unit 
 cells were relaxed.  In ferroelectric BTO, the cations and O of the alternating 
 BaO and TiO$_2$ layers are displaced against each other in the [001] direction.
 This leads to spontaneous polarization along [001]. 
 Here we model a dually polar ferroelectric. 
 If the BTO cations are placed above O in the supercell then the negative intralayer 
 displacements $\delta=(z_O-z_{cation}) <0$ form  the $\mathbf{P}$ state pointing parallel 
 to the surface normal ($P_\uparrow$) and, {\it vice versa}, the state $P_\downarrow$ 
 means that the $\delta >0$. Before relaxation, the $\delta$ values of 0.082 \AA\ and 
 0.086 \AA\ were chosen in the TiO$_2$ and BaO layers, respectively \cite{fechner1}. 
 The TiO$_2$-terminated type of the BTO interface was energetically preferred
 \cite{fechner1}. In this work, we substitute an interfacial Ti by Cr and added one or 
 two ML of iron on the CrO$_2$-terminated BTO(001).
 The Fe adatoms of the first ML relax atop oxygen \cite{Fechner2}, while the Fe 
 atoms of the second ML find their relaxed positions above the Ba and X=Cr sites.
 In Figure \ref{surface_cell}, we plot the side- and top view of relaxed Fe$_L$/CrO$_2$/BTO(001) 
 for the case of $L=2$. The positions of Fe above O, Ba and X are indicated by the corresponding 
 labels in the panel (b). 
 
 In this DFT based study we used the Vienna \emph{Ab initio} Simulation Package (VASP) 
 \cite{Kresse94,Kresse96,Hafner} within the local spin-density approximation.
 The electron-ion interactions were described by projector-augmented wave (PAW) pseudopotentials \cite{PAW}, 
 and the electronic wave functions were represented by plane waves with a cutoff energy of 650 eV. 
 For ionic relaxation the $8\times8\times4$ k-point  Monkhorst-Pack \cite{Monkhorst1976} mesh was used.
 The ionic relaxation was performed until the forces were less than $1\times10^{-3}$ eV/\AA.
 To calculate the electronic density of states (DOS) we used the $30\times30\times15$ k-point mesh.
 For each completely relaxed atomic configuration we performed the spin-polarized calculations 
 starting form the ferromagnetic (FM) or, alternatively, from the antiferromagnetic (AFM) configuration 
 in the Fe layers. The induced magnetization of the XO$_2$ interface was as well investigated.

\section{Results and discussion}

 Much effort has been recently put to show that the electric field-induced 
 reversal of ${\bf P}$ is able to vary the easy direction of magnetization in 
 magnetically soft Co$_{0.9}$Fe$_{0.1}$ \cite{Chu-NatMat2009} and Ni$_{0.78}$Fe$_{0.22}$
 permalloy \cite{Lebeugle-PRL2009} attached to thin film of multiferroic BiFeO$_3$ (BFO) or, 
 alternatively, to a single crystal of BiFeO$_3$. There is a problem, however, to form a ferroelectric
 single domain in the (001) plane of BFO. As a result, the magnetization of
 permalloy could not be completely switched. We suggest that BTO is a more promising
 material for switching $M$ by an electric field in the FM layer. 
 In our study of the 1-ML-thick Fe-electrode material deposited on BTO(001) we find that
 the two systems: Fe$_{L=1}$/TiO$_2$/BTO and Fe$_{L=1}$/CrO$_2$/BTO are 
 both ferromagnetically ordered, while
 the ME coupling coefficient increases from $\alpha =2.1\times10^{-10}$ G cm$^2$/V
 in Fe$_{L=1}$/TiO$_2$/BTO to the value of $7.2\times10^{-10}$ G cm$^2$/V at the CrO$_2$ interface.
 Eq.(1) was used to estimate $\alpha$.
 In the case of Fe bilayer, the magnetic order changes dramatically.
 The Fe$_{L=2}$/TiO$_2$/BTO system is almost zero-$M$ ferrimagnetic for the both ${\bf P}$ states. 
 Contrarily, Fe$_{L=2}$/CrO$_2$/BTO changes its magnetic order from AFM to FM 
 when the substrate polarization is switched from $P_\downarrow$ to $P_\uparrow$, resulting in a
 ME coupling coefficient of $\alpha =6\times10^{-8}$ G cm$^2$/V.
 Below we concentrate mainly on the case of $L=2$. 

\subsection{Structural relaxation}

\begin{figure}
  \includegraphics[width=1.0\linewidth]{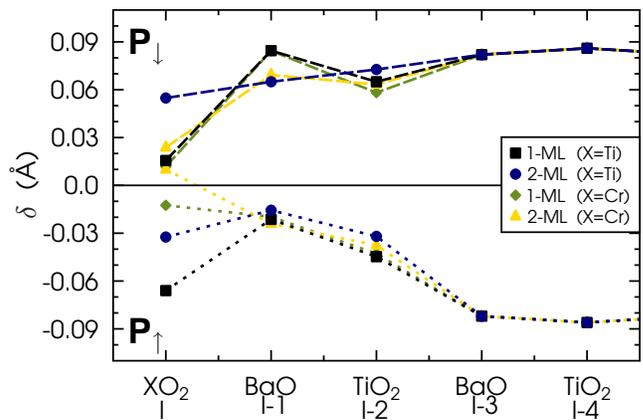}
  \caption{\label{displacement} Intralayer displacements $\delta=z_O-z_{cation}$ 
 (in \AA)  calculated for several top perovskite layers of Fe$_{L}$/XO$_2$/BaTiO$_3$(001) 
 ($L=1,2$, X=Ti,Cr and ${\bf P}= P_\downarrow, P_\uparrow$). 
  The interfacial layer and layers beneath are denoted by I, I-1, I-2, etc.}
\end{figure}

 Figure \ref{displacement} shows the perovskite intralayer displacement 
 between oxygen and cations along [001], $\delta  = z_{O} -z_{cation}$,
 obtained after relaxation of Fe$_{L}$/XO$_2$/BTO(001) ($L=1,2$, X=Ti,Cr and 
 ${\bf P}= P_\downarrow, P_\uparrow$). The interfacial layer and layers beneath 
 are denoted in Fig.~\ref{displacement} by I, I-1, I-2, etc.
 The asymmetry of $\delta$ seen between $P_\downarrow$ and $P_\uparrow$ for the 
 layers I, I-1 and I-2 as well as the magnitude of $\delta$, 
 which gradually decreases towards the interface, both mimic the effect of the
 depolarizing field and its screening.
 It should be noted that the state $P_\downarrow$ is energetically preferred 
 compared to $P_\uparrow$. For that reason the depolarization effect is 
 rather strong for $P_\uparrow$ as shown in Fig.\ref{displacement}. 
 For $P_\downarrow$, the value of $\delta$ is stable beneath the interface, 
 namely, between the layers I-1 and I-3 and, therefore, 
 the reduction of $\delta$ becomes 
 crucial at the interface only. It turns out that interfacial CrO$_2$ obeys marginal 
 $\delta$, which value decreases when the second Fe ML is added. For $P_\uparrow$, 
 the effect of X=Cr on $\delta$ is more pronounced. For instance, when $L=2$ 
 and ${\bf P}= P_{\uparrow}$ the presence of Cr changes the sign of $\delta$ 
 in layer I. 

\begin{figure}
  \includegraphics[width=1.0\linewidth]{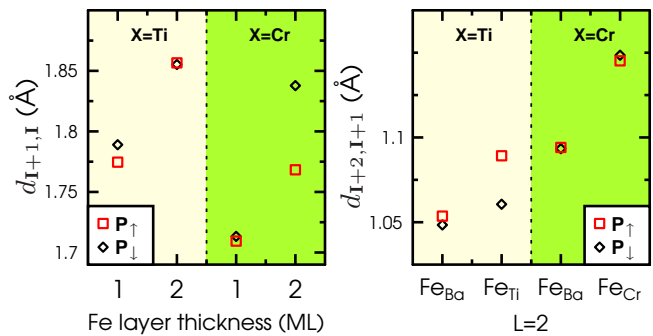}
  \caption{\label{fe_distances} Interlayer distances $d_{I+1,I}$ between the Fe 
 adlayer I+1 and interfacial O are shown versus the Fe thickness 
 $L=1,2$ in the left panel.
 For $L=2$, the distances $d_{I+2,I+1}$ between the topmost Fe$_{X}$/Fe$_{Ba}$ sites 
 and the Fe-(I+1) layer are shown in the right panel. Each system is dually polar.} 
\end{figure}
 In Figure \ref{fe_distances}, we plot the relaxed distances between interfacial Fe and O 
 atoms of XO$_2$ (X = Ti, Cr). It has been previously found from first principles that 
 the TiO$_2$ termination of BTO(001) is energetically preferred \cite{fechner1}. 
 When the first Fe ML is deposited on TiO$_2$/BTO(001) 
 the Fe atoms find their relaxed positions above O \cite{Fechner2} at the distance 
 $d_{I+1,I} \approx$1.78 \AA\, as shown in the left panel of Fig.~\ref{fe_distances}. 
 Thus, Fe and O form a strong and relatively short chemical bond at the  
 interface. Our calculations demonstrate that $d_{I+1,I}$ may increase by $\sim$5~\% 
 when the second Fe ML is added. The polarization reversal shows no effect on $d_{I+1,I}$.   
 For $L=1$ and the CrO$_2$-interface, we find that the corresponding $d_{I+1,I} \approx$1.7 \AA\
 is significantly reduced compared to the Fe/TiO$_2$/BTO systems. When the Fe-(I+2) layer 
 is added for X=Cr and $P_\uparrow$, the separation between Fe and O is increased to the 
 corresponding X=Ti value. For the opposite polarization $P_\downarrow$ and $L=2$, 
 a $\sim$5~\%-increase of $d_{I+1,I}$ was obtained. The latter result suggests a very
 promissing scenario of magnetoelectricity in the Fe$_L$/CrO$_2$/BTO system with $L=2$.
 Since the Fe-(I+2) atoms of the second layer are inevitably placed above the perovskite cations,
 the corresponding Fe$_{X}$ and Fe$_{Ba}$ sites are 
 nonequivalent as shown in Fig.~1. In Fig.~\ref{fe_distances}(b) we plot the relaxed 
 interlayer separation $d_{I+2,I+1}$ between the Fe layers I+2 and I+1 for the case 
 of $L=2$. In general, the presence 
 of Cr at the interface makes $d_{I+2,I+1}$ larger compared to the reference Fe$_L$/TiO$_2$/BTO
 system but, most importantly, $d_{I+2,I+1}$ is not changed upon {\bf P} reversal, exept for a 
 3~\%-increase at the Fe$_{Ti}$ site.    

\subsection{Electronic and magnetic properties}

\begin{figure}[bht]
\includegraphics[width=1.0\linewidth]{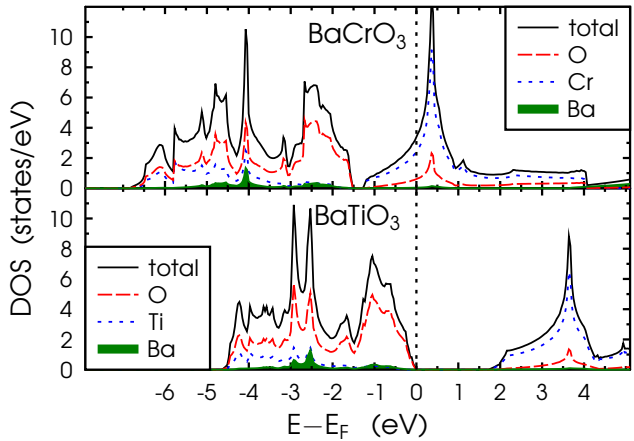}
\caption{\label{dos_BTO} The total and site-projected DOS of cubic BaTiO$_3$ and  
 hypothetic BaCrO$_3$ calculated using the same lattice parameter $a=3.943$ \AA .}
\end{figure}

 Fig. \ref{dos_BTO} shows the site-projected DOS of paraelectric cubic BaTiO$_3$ 
 together with the DOS of hypothetic cubic BaCrO$_3$. The two perovskites were 
 calculated using the same lattice parameter $a=3.943$~\AA . 
 For BTO we obtained an insulating band gap 
 of $\sim$2~eV, which is typically underestimated within the local density approximation. 
 The conduction band of BTO is formed mainly by the Ti 3$d$ states whereas the upper valence
 band is largerly composed by the O 2$p$ states. In BaCrO$_3$, the DOS is typically metallic while 
 the 3$d$ states of Cr dominate near the Fermi level, $E_F$. There is a marginal pseudogap 
 seen at -1.5 eV below $E_F$. Therefore, one can expect relatively strong metallization
 at the Fe/CrO$_2$ interface compared to Fe/TiO$_2$. 

\begin{table*}[htb]
  \caption{\label{atomar_moments} Local magnetic moments (in $\mu_B$) calculated for the two Fe 
  adlayers labeled by $I$+1 and $I$+2 and interfacial X (X = Cr, Ti) and O of 
  Fe$_{L}$/XO$_2$/BaTiO$_3$(001) ($L=1,2$). In the topmost Fe layer $I$+2, there are two nonequivalent 
 sites denoted as Fe$_{\text{Ba}}$ and Fe$_{\text{X}}$. The total magnetization $M_{tot}$ includes 
 the contributions from the interstitials. The energy difference between the AFM and FM
 configurations calculated for each system at ${\bf P}=$($P_{\uparrow},P_{\downarrow}$) is 
 shown in eV per cell.}
  \begin{ruledtabular}
    \begin{tabular}{ll|cc|cc|cc|cc}
      Site&Layer&  \multicolumn{2}{c|}{(Fe$_2$)$_{L=1}$/CrO$_2$/BTO} &  
 \multicolumn{2}{c|}{(Fe$_2$)$_{L=2}$/CrO$_2$/BTO} & \multicolumn{2}{c|}{(Fe$_2$)$_{L=1}$/TiO$_2$/BTO} & 
 \multicolumn{2}{c}{(Fe$_2$)$_{L=2}$/TiO$_2$/BTO}\\  
      & & $P_\uparrow$ &$P_\downarrow$&$P_\uparrow$&$P_\downarrow$& 
 $P_\uparrow$ &$P_\downarrow$&$P_\uparrow$&$P_\downarrow$\\
      \hline
    
      Fe$_{\text{Ba}}$  &(I+2) & ---     & ---     & $+2.00$ & $+2.24$ & ---     & ---     & $+2.41$ & $+2.36$ \\
      Fe$_{\text{X}}$   &(I+2) & ---     & ---     & $+2.41$ & $-2.61$ & ---     & ---     & $-2.46$ & $-2.36$ \\
      Fe$_{\text{O}}$   &(I+1) & $+2.72$ & $+2.75$ & $+0.86$ & $+0.44$ & $+2.83$ & $+2.81$ & $-0.03$ & $+0.00$ \\
      X                 &(I)   & $-2.10$ & $-2.00$ & $+1.79$ & $-0.11$ & $-0.30$ & $-0.22$ & $ 0.00$ & $+0.01$ \\
      O                 &(I)   & $+0.10$ & $+0.11$ & $-0.03$ & $ 0.00$ & $+0.09$ & $+0.08$ & $-0.01$ & $-0.01$ \\
      \hline
      \multicolumn{2}{l|}{$E_{AFM}-E_{FM}$(eV)}& $+0.65$ & $+0.70$ & $+0.01$ & $-0.02$ & $+0.69$ & $+0.75$ & $-0.12$ & $-0.12$ \\   
      \multicolumn{2}{l|}{$M_{tot}$ ($\mu_B$)} & $+3.86$ & $+3.96$ & $+8.28$ & $+0.28$ & $+5.87$ & $+5.84$ & $+0.02$ & $-0.02$ \\
      \multicolumn{2}{l|}{$\alpha$ (G cm$^2$/V)} & \multicolumn{2}{c|}{$7.20\times10^{-10}$} & \multicolumn{2}{c|}{$5.99\times10^{-8}$} & \multicolumn{2}{c|}{$2.08\times10^{-10}$} & \multicolumn{2}{c}{3.05$\times10^{-10}$} 
    \end{tabular}
  \end{ruledtabular}
\end{table*}

 In Fe$_{L=1}$/TiO$_2$/BTO the FM order is energetically favorable against the AFM
 solution by 0.7~eV/cell (0.75~eV/cell) for $P_{\uparrow}$ ($P_{\downarrow}$).
 Here, the Fe and O magnetic moments are aligned parallelly whereas the Ti magnetic moment, 
 originating from hybridization of the Ti $3d$ and Fe $3d$ minority states \cite{Duan:2006p278}, 
 is antiparallelly aligned. All magnetic moments of the system are collected in 
 Table \ref{atomar_moments}. 
 The polarization reversal from $P_\downarrow$ to $P_\uparrow$ yields the magnetization 
 change $|\Delta M|=0.028$~$\mu_B$/cell which formally results in the ME coupling of 
 2.1$\times10^{-10}$~G cm$^2$/V. 
 When Cr substitutes Ti at the interface, the lowest-energy configuration
 remains ferromagnetic. However, the negative magnetic moment of $\sim$2~${\mu}_B$, 
 induced on Cr, is much larger than $m_{Ti}$. For interfacial oxygen the calculated magnetic
 moment is about 0.1~$\mu_B$. This value as well as $m_{Cr}$ are in a good agreement 
 with the experimental data of bulk CrO$_2$ \cite{Huang2002}.    
 Due to the large and negative Cr magnetic moment, the total magnetization of the system
 Fe$_{L=1}$/CrO$_2$/BTO is reduced by $\approx2$ $\mu_B$ in comparison to that of 
 Fe$_{L=1}$/TiO$_2$/BTO. Although $m_{Cr}$ is moderately changed by ${\bf P}$ reversal
 the corresponding $|\Delta M|$ results in $\alpha=$ 7.2$\times10^{-10}$ G cm$^2$/V,
 which is three times larger than the ME effect of Fe$_{L=1}$/TiO$_2$/BTO. 

\begin{figure}[htb]
\includegraphics[width=1.0\linewidth]{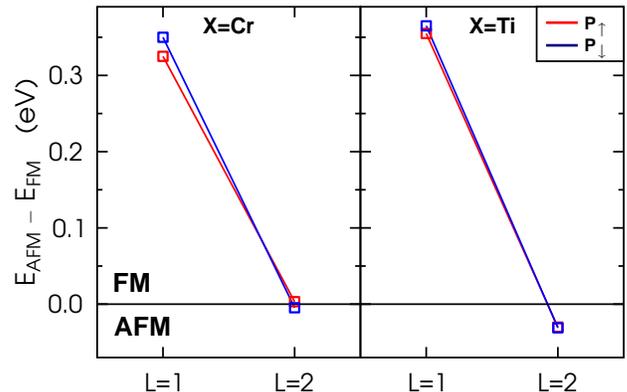}
\caption{\label{deltaE} Energy difference $\Delta E = E_{AFM} - E_{FM}$ between the 
 AFM and FM configurations of Fe$_{L}$/XO$_2$/BTO (X=Cr,Ti and $L=1,2$) is
 normalized per Fe atom.}
\end{figure}

 The second Fe ML deposited on the TiO$_2$-terminated BTO(001) interface 
 causes a specific case.
 There are two inequivalent I+2 sites situated atop Ba and Ti,
 respectively, which are labelled by Fe$_{Ba}$ and Fe$_{X}$ in Fig.~1.
 The different magnetic moments reflect the neighbourhood of these atoms such as 
 their atomic volumes and hybridization of the electronic states.
 Let us consider, first, the case of X=Ti. 
 The value of $m_{Fe}$ in the layer I+1 is almost quenched while the two sizable moments 
 in the surface layer I+2 are antiparallelly aligned. This results in $M \to 0$ for 
 Fe$_{L=2}$/TiO$_2$/BTO(001). In the case of the Fe bilayer on the CrO$_2$-terminated BTO,  
 the lowest-energy configuration is antiferromagnetic for $P_{\downarrow}$ and
 becomes ferromagnetic for $P_{\uparrow}$. For this polarization
 the Fe magnetic moments in the layer I+1 are far below their bulk value but
 the two Fe-(I+2) magnetic moments, which are ferromagnetically aligned 
 to each other, contribute significantly to the total $M$.
 We estimate that the total magnetic moment of the system changes  
 from $M <$0.3~$\mu_B$ to $>$8~$\mu_B$ per unit cell area upon polarization reversal.
 Thus, the polarization reversal produces for X=Cr the effect of switchable magnetization. 
 In Fig.~\ref{deltaE}, the difference in energy, $\Delta E = E_{AFM} - E_{FM}$, calculated 
 between the AFM and FM configurations and normalized per Fe atom, is plotted. 
 For X=Cr the 2-ML-thick Fe film represents a specific case of a magnetically soft system
 at fixed ${\bf P}$. Nevertheless, any magnetic switch upon ${\bf P}$ 
 reversal requires an energy which exceeds the coercive field value of BTO.

\begin{figure}[hbt]
 \centering
 \includegraphics[width=1.0\linewidth]{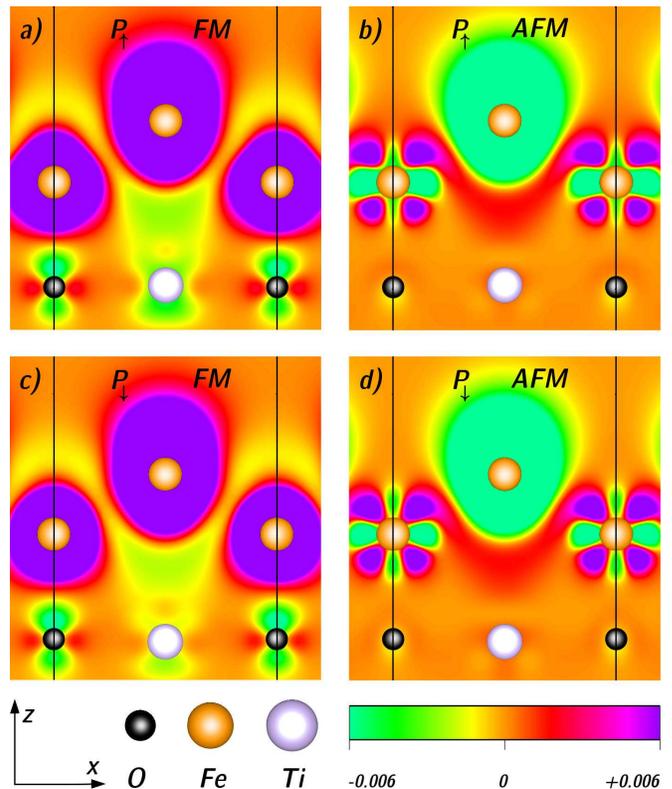}
 \caption{\label{mag_dens_Ti} Spin density imbalance $n^+(\mathbf{r})-n^-(\mathbf{r})$
 (in e/\AA$^{-3}$) within the (100) plane cutting through the Ti atoms of the
 Fe$_{L=2}$/TiO$_2$/BaTiO$_3$(001) slab. The black vertical lines represent
 the unit cell boundary. The two top (bottom) panels show the polarization
 $P_\uparrow$ ($P_\downarrow$).
 The panels (a) and (c) illustrate the FM ordering obtained for $P_\uparrow$
 and $P_\downarrow$, respectively, while the energetically preferable and
 nearly AFM configurations are shown in the panels (b) and (d).}
\end{figure}

\begin{figure}[thb]
\centering
\includegraphics[width=1.0\linewidth]{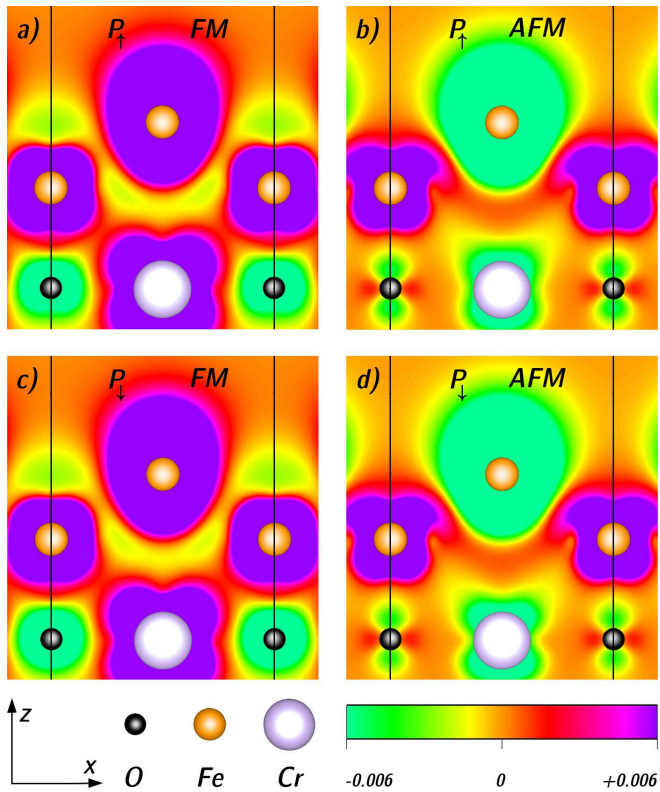}
 \caption{\label{mag_dens_Cr} Spin density imbalance (in e/\AA$^{-3}$) within the (100) 
 plane cutting through the Cr atoms of Fe$_{L=2}$/CrO$_2$/BaTiO$_3$(001).
 The two top (bottom) panels show the $P_\uparrow$ ($P_\downarrow$) states while 
 the left (right) panels illustrate the FM (AFM) ordering. For $P_\uparrow$ 
 ($P_\downarrow$), the lowest energy solution is the FM (AFM) configuration 
 shown in `a' (`d').}
\end{figure}

 To illustrate the interface ME coupling mechanism, we plot in Fig.~\ref{mag_dens_Ti}
 and Fig.~\ref{mag_dens_Cr} the spin density imbalance, ($n^+(\mathbf{r})-n^-(\mathbf{r})$), 
 obtained under $P$-reversal near the interface of Fe$_{L=2}$/TiO$_2$/BTO 
 and Fe$_{L=2}$/CrO$_2$/BTO, respectively.
 The (100) plane cutting through the X and O interfacial sites shows where the largest 
 changes of the spin density occur and, hence, from where the ME effect arises.
 Each of the four panels of Fig.~\ref{mag_dens_Ti}--\ref{mag_dens_Cr} shows the local
 magnetization density calculated at fixed ${\bf P} =$(${P_\downarrow},{P_\uparrow}$).
 These are shown for the two possible magnetic configurations which are either FM or AFM.
 For X=Ti, both the $P_\uparrow$- and $P_\downarrow$-poled states are antiferromagnetically ordered, 
 as shown in the panels (b) and (d) of Fig.~\ref{mag_dens_Ti}. The two results are similar to each other.
 The largest negatively charged areas are seen around Fe$_X$-(I+2) while the $n^+$- charged
 areas around the second Fe site of this layer are not shown in Fig.~\ref{mag_dens_Ti}. 
 All other sites of Fe$_{L=2}$/TiO$_2$/BTO including Fe-(I+1) indicate very small magnetic moments.
 Inspecting the spin density imbalance seen in Fig.\ref{mag_dens_Cr}(a) and Fig.\ref{mag_dens_Cr}(d)
 for the two energetically preferred but oppositely poled configurations of X=Cr,
 we find many differences in the magnetic structure. The panel (a) shows the ferromagnetically
 ordered state $P_\uparrow$ where the Fe and Cr atoms form rather spacious regions of positive spin density
 $n^+$ while $n^-$ can be spotted around O, in the Fe interstitials and regions towards the surface.
 In the case of $P_\downarrow$, the energetically favorable AFM configuration, shown in the panel (d),
 is similar to that of Fe$_{L=2}$/TiO$_2$/BTO. Here, the large areas around Fe$_X$-(I+2) and also
 around interfacial Cr are negatively charged. Besides, the p$_z$-orbitals of interfacial O show their
 negative spin population resulting from hybridization with the 3$d$ states of Fe-({\bf I+1}) whereas
 the O p$_x$ and p$_y$ orbitals, which form the bonds with the Cr 3d states, contribute to $n^+$.
 Regarding the Fe-({\bf I+1}) atoms of Fe$_{L=2}$/CrO$_2$/BTO, Fig.\ref{mag_dens_Cr}(d) shows that
 they contribute to $n^+$ contrarily to the case of X=Ti.


\begin{figure}[bht]
\includegraphics[width=1.0\linewidth]{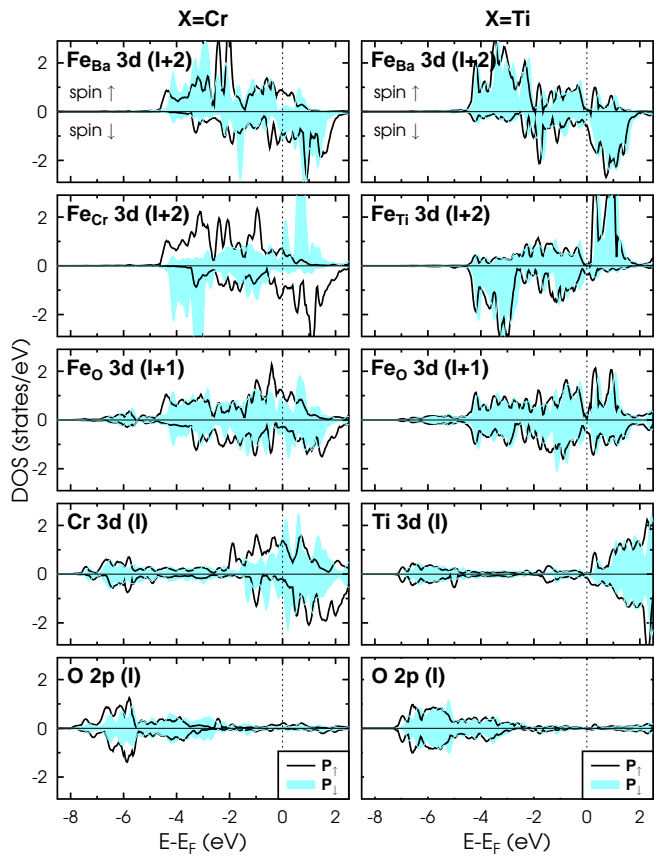}
\caption{\label{surface_dos} Spin-polarized and site-projected DOS calculated for the
 metal 3$d$- and O 2$p$-states near the inerface of (Fe$_2$)$_{L=2}$/CrO$_2$/BaTiO$_3$(001)
 (left) and (Fe$_2$)$_{L=2}$/TiO$_2$/BaTiO$_3$(001) (right). The two upper panels show the Fe
 3$d$ DOS of  the surface layer I+2 while the Fe-I+1 DOS are shown in the 
 middle panels. The 3$d$-DOS of inerfacial cations Cr/Ti and oxygen 2$p$-DOS
 are plotted in the two lower panels. Solid lines and shaded areas represent the DOS for
 $P_{\uparrow}$ and $P_{\downarrow}$, respectively.}
\end{figure}

 The site-projected and spin-resolved DOS calculated for Fe$_{L=2}$/XO$_2$/BTO are 
 plotted in the two panels of 
 Fig. \ref{surface_dos}. For each system, the solid (shaded) lines represent 
 the DOS curves in the $P_{\uparrow}$ ($P_{\downarrow}$) state. The energetically preferable 
 magnetic configurations are shown only in Fig. \ref{surface_dos} for each direction of $\bf P$. 
 In general, the DOS of the  interfacial XO$_2$ layer is metallic for both systems. 
 For $L=2$ and X=Cr, however, the Cr 3$d$--DOS indicates relatively strong spin polarization 
 at the Fermi level. This is not surprising since the DOS of hypothetical BaCrO$_3$ shows 
 similar behavior, as shown in Fig.~4. When X=Ti, there is some insignificant 
 presence of the Ti 3$d$ states in the BTO band gap below $E_F$, which entirely results 
 from the hybridization with the Fe $3d$ states of the layer I+1.  
 Another major difference in the DOS seen in Fig. \ref{surface_dos} for $L=2$
 comes from the magnetic ordering of Fe$_X$. For X=Ti the two Fe atoms in the topmost layer 
 I+2 are coupled antiferrimagnetically while the corresponding DOS curves show minor changes 
 upon {\bf P} reversal. When X=Cr the polarization reversal from $P_{\downarrow}$ to the state 
 $P_{\uparrow}$ supports (i) the ferromagnetic order in the layer I+2, (ii) the relatively large
 magnetic moment $m$(Fe$_O$)$\sim$0.9~${\mu}_B$ in the layer I+1 and (iii) the
 $\sim$2-${\mu}_B$ change of $m_{Cr}$ which is aligned parallelly to the Fe magnetic moments. 

\begin{figure}[thb]
\centering
\includegraphics[width=1.0\linewidth]{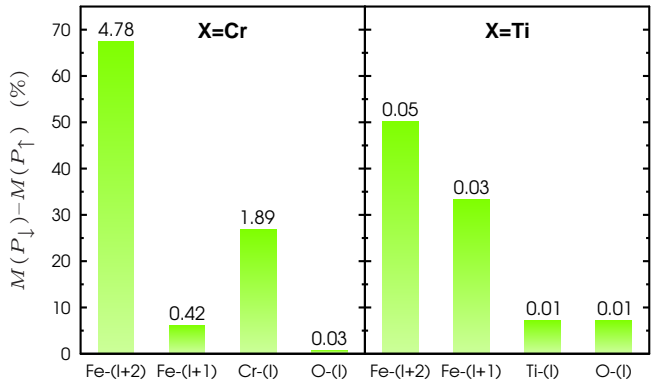}
 \caption{\label{mag_contribut} 
 Relative contributions (in \%) of each magnetic species of Fe$_{L=2}$/XO$_2$/BTO 
 (X = Ti, Cr) to the magnetization change, $\Delta M$, induced by polarization reversal.
 The contributions from the two Fe-(I+2) and two Fe-(I+1) atoms, one interfacial X and two O atoms 
 were considered. The absolute values of $\Delta M$ are given (in ${\mu}_B$) above each bar.}
\end{figure}

 In Fig.~\ref{mag_contribut}, we plot the relative (in \%) and absolute contributions 
 (in ${\mu}_B$) to $\Delta M = M(P_{\downarrow})-M(P_{\uparrow})$ coming from each 
 magnetic species of Fe$_{L=2}$/XO$_2$/BTO. 
 For the two biferroic interfaces studied here, the largest ${\bf P}$-induced change of 
 ${\bf M}$ comes from the Fe-(I+2) atoms. For X=Cr, however, the absolute value of $\Delta M$ 
 approaches $\sim$7~${\mu}_B$ per unit cell. As result, the corresponding ME coupling 
 coefficient increases significantly compared to that of X=Ti. We demonstrate that the case 
 of $L=2$ and X=Cr stabilizes the FM ordering in the system with {\bf P} pointing upwards. 
 Surprisingly, this is completely due to rather modest 5~\% decrease of $d_{I+1,I}$  
 under {\bf P} reversal, as shown in Fig.~3. With decreasing $d_{I+1,I}$ 
 above CrO$_2$, the FM order is developing in the system. More precisely, when the Fe-(I+1) 
 magnetic moment becomes larger ferromagnetism is stabilized in layer (I+2).   
 In the case of X=Ti, the interlayer separations $d_{I+2,I+1}$ and $d_{I+1,I}$ are almost 
 the same upon the {\bf P} reversal that prevents any crucial spin reorientation in the 
 topmost Fe ML.   

\section{Conclusions}

 In summary, we present an {\it ab initio} study of the effect of interfacial Cr on the 
 strength of magnetoelectric coupling seen at the interface of multiferroic 
 Fe$_L$/CrO$_2$/BaTiO$_3$(001), with the Fe thickness $L\le$2 monolayers.
 We predict that a CrO$_2$-terminated interface instead of TiO$_2$ may 
 significantly enhance magnetoelectricity in the system. The most attractive scenario 
 is, however, obtainded for the Fe bilayer where the magnetic order changes from
 nearly zero-${\bf M}$ ferrimagnetic to ferromagnetic upon  
 polarization reversal in ferroelectric BaTiO$_3$(001).

\section{Acknowledgments}

 This work was supported by the Collaborative Research Network SFB 762, 'Functionality of Oxidic 
 Interfaces'. M. Fechner is a member of the International Max Planck Research School for Science and 
 Technology of Nanostructures.


\end{document}